\documentclass[twocolumn,prb,showpacs,floatfix]{revtex4}
\usepackage{graphicx}
\usepackage{dcolumn}
\usepackage{bm}
\begin{document}
\title{Tomonaga--Luttinger parameters and spin excitations 
\\in the dimerized extended Hubbard model}
\author{Satoshi Ejima$^1$, Florian Gebhard$^1$, and Satoshi Nishimoto$^2$}
\affiliation{$^1$Fachbereich Physik, Philipps-Universit\"at Marburg, 
D-35032 Marburg, Germany\\
$^2$Max-Planck-Institut f\"ur Physik komplexer Systeme, 
D-01187 Dresden, Germany} 
\date{\today}
\begin{abstract}%
We study the one-dimensional extended 
Hubbard model with alternating size of the hopping integrals 
using the density-matrix renormalization group method. 
We calculate the spin gap, the Tomonaga--Luttinger
parameter, and the charge-density-wave order parameter 
for various dimerizations, interaction strengths, and band fillings. 
At half band-filling the spin and charge excitations are gapped but these gaps 
disappear for infinitesimal hole doping. 
At quarter filling, the Umklapp scattering in the half-filled lower
Peierls band generates a gap for the charge excitations but
the gapless spin excitations can be described in terms of an effective
antiferromagnetic Heisenberg model. 
Beyond a critical strength for the nearest-neighbor interaction,
the dimerized extended Hubbard model at quarter filling 
develops a charge-density-wave ground state.
The dimerization and the nearest-neighbor Coulomb interaction
strongly reduce the Tomonaga--Luttinger parameter from its value for
the bare Hubbard model. We discuss the relevance of our findings
for the Bechgaard salts.
\end{abstract}
\maketitle

\section{INTRODUCTION}

The Bechgaard salts are organic conductors which have attracted 
much interest over the last thirty years~\cite{ishiguro,lang}. 
Upon variation of the pressure, the temperature, and
the anion~$X$ in (TMTSF)$_2$X and (TMTTF)$_2$X, 
these compounds exhibit a rich phase diagram, e.g., 
a superconducting phase is found to lie in-between a paramagnetic 
metallic phase and a spin-density-wave phase. The systems can 
be regarded as quasi one-dimensional due to the 
strong anisotropy of the transport along the three crystalline axes. 
Recent experiments~\cite{vescoli} support the view that 
the metallic phase can be characterized as a Tomonaga--Luttinger 
liquid at temperatures $T>100$ K. Indeed, signatures of the
Tomonaga--Luttinger liquid are 
the reduced density of states at the Fermi energy 
as seen in angle-resolved photoemission spectroscopy~\cite{dardel,zwick}, 
the negative temperature dependence of the c-axis resistivity~\cite{moser},
the scaling behavior of high-energy range of the optical 
conductivity~\cite{dressel}, the power-law temperature dependence 
of the Hall coefficient~\cite{mihaly,moser2},
and the empirical relationship $(T_1T)^{-1} \propto \chi_s^2(T)$ 
between the measured spin relaxation rate and the magnetic susceptibility 
in nuclear magnetic resonance measurements~\cite{bourbonnais,wzietek}.
Moreover, distinctly different thermal conductivities 
for the charge and spin excitations have been reported which provide
evidence for spin-charge separation~\cite{lorenz}.

All correlation functions in the Tomonaga--Luttinger liquid display
a power-law behavior with unusual, inter\-action-dependent coefficients.
Many of them are simple functions of the so-called Tomonaga--Luttinger 
parameter~$K_{\rho}$. Most experiments give 
$K_{\rho} \approx 0.2$ for the Bechgaard salts.
The single-band Hubbard model in which
spin-1/2 electrons move on a chain
and interact only locally is one of the best studied Hamiltonians
for correlated lattice electrons. However,
the model gives $K_{\rho} \geq 0.5$ 
for all interaction strengths which shows that the long-range parts of
the Coulomb interaction must be taken into account for
a proper description of the Bechgaard salts.
In the extended Hubbard model the long-range parts of the Coulomb interaction
are mimicked by a nearest-neighbor term~\cite{nishim1,mila,penc}. 

Other factors may also play an important role.
For instance, the stacks of TMTTF and TMTSF molecules form dimerized chains
and the alternation of the electron transfer-matrix elements 
along the chain must be considered.
Therefore, in this work we study the one-dimensional extended Hubbard model 
with alternating hopping amplitudes, i.e., 
the one-dimensional dimerized extended Hubbard model
as the minimal one-dimensional, purely electronic model 
for the electronic excitations in the Bechgaard salts.
The relevant bands in the TMTSF and TMTTF salts are filled with three
electrons so that the system is quarter-filled in hole notation,
and we use the hole picture in the following. 

There are few systematic studies of the dimerized extended
Hubbard model in the literature. Therefore, we investigate
the model for various band fillings, with an emphasis
on the vicinity of the commensurate fillings. In this way,
our principle investigation of correlated electrons in 
quasi one-dimensional dimerized systems could be relevant also
for other materials, 
e.g., for the inorganic spin-Peierls system CuGeO$_3$~\cite{hase}.

In our work we apply the density-matrix renormalization group (DMRG) 
method which is one of the most reliable numerical methods 
to study the 
low-energy properties of one-dimensional correlated electron systems.
Where applicable, we compare our results to the predictions 
from field theory and effective single-band Hubbard models.

Our paper is organized as follows. In Sec.~\ref{sec:modelmethod}
we define the dimerized extended Hubbard model and introduce
the physical quantities of interest, namely, the spin gap, 
the charge-density-wave (CDW) order parameter, and the Tomonaga--Luttinger parameter.
In Sec.~\ref{sec:results}, we separately present our DMRG results for
the dimerized Hubbard model with and without the nearest-neighbor interaction,
and discuss the experimental relevance of our investigations.
We close with a short summary in Sect.~\ref{sec:sum}.

\section{MODEL AND METHOD}
\label{sec:modelmethod}

\subsection{Hamiltonian}

In order to model the Bechgaard salts, we focus on
the transport of a chain of stacked molecules
and regard a single TMTTF or TMTSF molecule as a site. 
The chain has a geometrical 
(Peierls) modulation. Besides the intra-molecular Coulomb interaction, 
we should take into account a nearest-neighbor Coulomb repulsion because
of the fairly short inter-molecular distance. 
Thus, our model Hamiltonian of choice is the one-dimensional
dimerized extended Hubbard model for spin-1/2 electrons
on~$L$ lattice sites
\begin{eqnarray}
\nonumber
\hat{H} \!&=&\! -t_1 \sum_{l,{\rm odd}} 
(\hat{c}^{\dagger}_{l+1\sigma} \hat{c}_{l\sigma} 
+ {\rm h.c.}) 
- t_2 \sum_{l,{\rm even}} (\hat{c}^{\dagger}_{l+1\sigma} \hat{c}_{l\sigma} 
+ {\rm h.c.}) \\
&&+ U \sum_l \hat{n}_{l\uparrow} \hat{n}_{l\downarrow} 
+ V \sum_l (\hat{n}_l - n)(\hat{n}_{l+1} - n) \; ,
\label{ham}
\end{eqnarray}
where $\hat{c}^{\dagger}_{l\sigma}$ ($\hat{c}_{l\sigma}$) 
is the creation (annihilation) operator of an electron 
with spin $\sigma=\uparrow,\downarrow$ at site~$l$, 
$\hat{n}_{l\sigma}=\hat{c}^{\dagger}_{l\sigma}\hat{c}_{l\sigma}$
is the number operator, 
and  $\hat{n}_l=\hat{n}_{l\uparrow}+\hat{n}_{l\downarrow}$. 
The total number of electrons is $N=N_{\uparrow}+N_{\downarrow}$,
and $n=N/L$ is the average number of electrons per lattice site.
The electron transfer matrix elements $t_1$ and $t_2< t_1$ model
the dimerization of the chain, $U$ is the strength of the Hubbard interaction, 
and $V$ parametrizes the nearest-neighbor Coulomb repulsion. 
We call a pair of sites which is connected by the hoping amplitude~$t_1$
a `dimer'. 

\begin{figure}[ht]
\includegraphics[width=7.5cm]{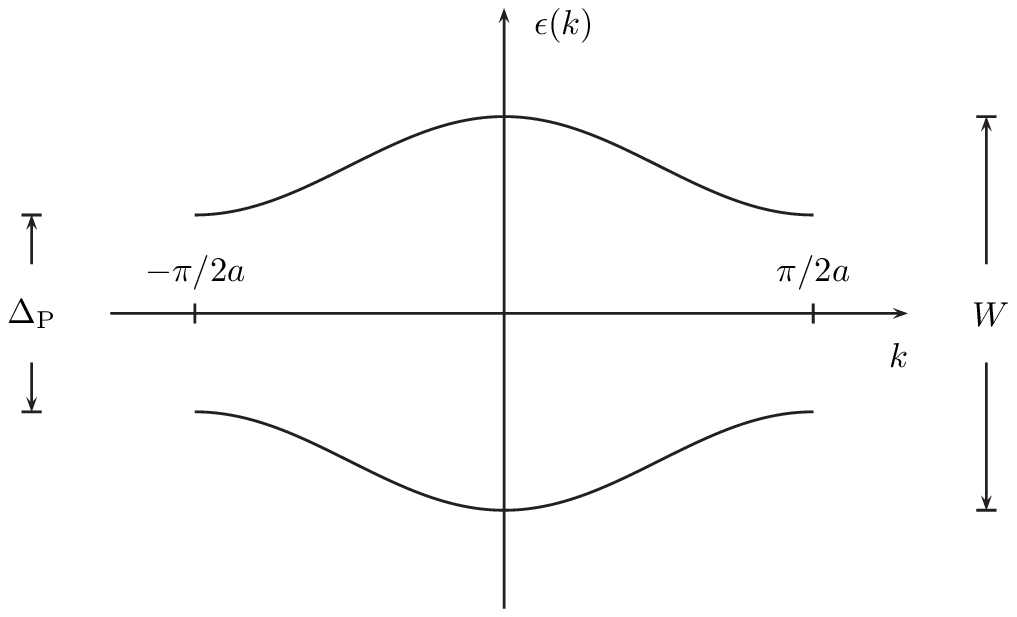}
\caption{Band structure for non-interacting electrons in a dimerized chain.}
    \label{Fig:barebands}
\end{figure}

The dimerization splits the tight-binding cosine band into a bonding band
(`lower Peierls band') and an anti-bonding band (`upper Peierls band').
The bare band structure is shown in Fig.~\ref{Fig:barebands}.
The dispersion relation of the two Peierls bands is given by
\begin{equation}
\epsilon_{1,2}(k) = \pm \sqrt{t_1^2+t_2^2+2t_1t_2\cos k} \quad \hbox{for}
\quad |k|\leq \frac{\pi}{2a}\; ,
\end{equation}
where $a$ is the lattice spacing which we set to unity in the following.
The gap between the two Peierls bands is $\Delta_{\rm P}=2(t_1-t_2)$. 
The total band width is $W=2(t_1+t_2)$.
In the absence of a dimerization, for $t_1=t_2=t$, 
we recover the band structure of the tight-binding model 
in the reduced zone scheme.

\subsection{Physical quantities}

In this work we employ the DMRG method 
which provides very accurate data for ground-state properties
of one-dimensional correlated electron systems; for a review, 
see~[\onlinecite{DMRGbook,Schollwoeck}].
We use the DMRG to calculate
the spin gap $\Delta_s$, the CDW 
order parameter $\chi$, and the Tomonaga--Luttinger parameter $K_\rho$.
To this end, we consider a chain with 
$L/2$ dimers with $L/2$ even for a two-band system.
We study chains with up to $320$ sites 
and open-end boundary conditions.
We keep up to $m=3600$ density-matrix eigenstates 
in the DMRG procedure and extrapolate the calculated quantities 
to the limit $m \to \infty$. In this way, the maximum 
error in the ground-state energy is below $10^{-6}t_1$. 
Lastly, we extrapolate our finite-size results to the thermodynamic
limit, $L\to\infty$.

The spin gap is defined by
\begin{eqnarray}
\nonumber
\Delta_s &=& \lim_{L \to \infty} \Delta_s (L) \; , \\
 \Delta_s (L) &=& E_0(L, N_{\uparrow}+1, N_{\downarrow}-1) 
- E_0(L, N_\uparrow, N_\downarrow) \; , 
\label{define-spingap}
\end{eqnarray}
where $E_0(L, N_\uparrow, N_\downarrow)$ 
is the ground state energy of a system of length $L$ 
with $N_{\uparrow}$ up-spin and $N_{\downarrow}$ down-spin electrons. 

Later in this work, we shall focus on the CDW ground state
of our model~(\ref{ham}) at quarter band filling. For large enough 
nearest-neighbor repulsion~$V$ 
we expect a CDW 
with a wave vector $Q_{\rm CDW}=4k_{\rm F}$. Here, $k_{\rm F}=\pi n/2$
is the Fermi wave number. At quarter band filling, $n=1/2$, 
we have $k_{\rm F}=\pi/4$ which corresponds to 
a half-filled lower Peierls band.

The order parameter for the $4k_{\rm F}$-CDW phase 
is defined by
\begin{eqnarray}
\chi &=& \lim_{L \to \infty} \chi(L)\; , \\
\chi(L) &=& \left| \frac{1}{r+2}\sum_{l=(L-r)/2}^{(L+r)/2+1} 
(-1)^l \left\langle \hat{n}_l \right\rangle \right| \; .
\label{orderparameter}
\end{eqnarray}
In~(\ref{orderparameter}) the summation over the lattice sites~$l$ 
is restricted to a region~$r$ around the central site of the chain in order to 
reduce the edge effects. We set $r=2$ for a systematic 
extrapolation to the thermodynamic limit. Of course, the extrapolated results 
should be independent of the choice of the range~$r$. 
On finite lattices and for open-end boundary conditions, the Friedel oscillations 
from the edges result in a finite value for $\chi(L)$,
and a well-controlled finite-size extrapolation is mandatory.

For the calculation of the Tomonaga--Luttinger parameter $K_\rho$ we use
a new method which we proposed recently~\cite{ejima}.
The Tomonaga--Luttinger parameter $K_\rho$ determines the
long-range decay of the density-density correlation function
in the metallic Tomonaga--Luttinger liquid ground state.
It is defined by the ground-state expectation value
\begin{equation}
C^{\rm NN}(r) 
= \frac{1}{L} \sum_{l=1}^{L} 
\langle \hat{n}_{l+r} \hat{n}_{l} \rangle
- \langle \hat{n}_{l+r} \rangle 
\langle \hat{n}_{l} \rangle \; .
\label{CNNr}
\end{equation}
Using conformal field theory it can be shown~\cite{Fra90,Sch90} that
the asymptotic behavior for $1\ll r\ll L$ is given by
\begin{equation}
C^{\rm NN}(r) \sim
-\frac{K_{\rho}}{(\pi r)^2}
+\frac{A\cos(2k_{\rm F}r)}{r^{1+K_{\rho}}}\ln^{-3/2}(r)
+\cdots\; ,
\label{eqn:den1}			 
\end{equation}
where $A$~is a constant. In previous 
approaches~\cite{dzierzawa,noack,daul,clay2}, 
$K_{\rho}$ was extracted from the Fourier transformation 
of~$C^{\rm NN}(r)$ but in a real-space DMRG approach
the accuracy of the correlation function becomes increasingly worse 
as the distance~$r$ increases.
In~Ref.~[\onlinecite{ejima}] 
we calculated the density-density correlation function
directly in Fourier space. We address 
\begin{eqnarray}
N(q)=\frac{2}{L} \left\langle \hat{n}(q)\hat{n}(-q) \right\rangle \; ,
\label{Nq}
\end{eqnarray}
where $\hat{n}(q)$ is given by
\begin{eqnarray}
n(q)=\sum_{l,{\rm odd}} e^{{\rm i}(q/2)(l+1/2-r_c)}
(\hat{c}^{\dagger}_{l\sigma} \hat{c}_{l\sigma} 
+ \hat{c}^{\dagger}_{l+1\sigma} \hat{c}_{l+1\sigma})\; .
\label{nq}
\end{eqnarray}
Here, $r_c = (L + 1)/2$ denotes the central position of the chain. 
The derivative of $N(q)$ at $q = 0$ directly
gives the Tomonaga--Luttinger parameter. In practice, we
obtain it from
\begin{eqnarray}
K_{\rho}&=& \lim_{L \to \infty} K_{\rho}(L) \nonumber \; ,\\
K_{\rho}(L)&=&\frac{L}{4} N\left(\frac{4\pi}{L}\right) \; . 
\label{KrhoD}
\end{eqnarray}
For a precise calculation of $K_{\rho}$ is important to target not only 
the ground state $| \Phi_0 \rangle$ but also the state 
$| \Psi_q \rangle = \hat{n}(-q) | \Psi_0 \rangle$ in the DMRG procedure;
see~Ref.~[\onlinecite{ejima}] for further details.

The Tomonaga--Luttinger parameter is well defined only for the metallic
Tomonaga--Luttinger liquid. Later we shall investigate $K_{\rho}$
for insulators which are infinitesimally doped away from
their commensurate doping $n_{c}$. In these cases we give
\begin{eqnarray}
K_{\rho}(n \to n_c^{\pm})=\lim_{L \to \infty} 
K_{\rho}\left(n=n_c \pm \frac{2}{L}\right)\; .
\label{Krhoquarter}
\end{eqnarray}
This approach is very successful for the single-band Hubbard model,
as demonstrated in~Ref.~[\onlinecite{ejima}].

\subsection{Effective models}

For not too small dimerizations, $t_2/t_1\lesssim 0.9$,
we can map the dimerized extended Hubbard model to an effective single-band
extended Hubbard model~\cite{nishim1}. 
The upper Peierls band can be integrated out and
we are left with a Hubbard chain with $L/2$ dimer sites~$l_{\rm d}$ 
with effective parameters,
\begin{eqnarray}
\hat{H}_{\rm eff} & = & t_{\rm eff} 
\sum_{l_{\rm d}} 
(\hat{c}^{\dagger}_{l_{\rm d}+1\sigma} \hat{c}_{l_{\rm d}\sigma} 
+ {\rm h.c.}) 
+ U_{\rm eff} 
\sum_{l_{\rm d}} \hat{n}_{l_{\rm d}\uparrow} 
\hat{n}_{l_{\rm d}\downarrow}\nonumber \label{hameffext}  \\
&& +  V_{\rm eff} \sum_{l_{\rm d}} (\hat{n}_{l_{\rm d}} - 1)
(\hat{n}_{l_{\rm d}+1} - 1) \; ,\\
t_{\rm eff} &=& \frac{t_2}{2} \label{teffext} \; , \\
U_{\rm eff} &=& 2t_1-\frac{\sqrt{(U-V)^2+16t_1^2}-(U+V)}{2}\; , 
\label{effUext}\\
V_{\rm eff} &=& \frac{V}{4}\; . \label{effVext}
\end{eqnarray}
The band filling is $n_{\rm eff}=2n$ so that $k_{\rm F, eff}=\pi n$ and
$v_{\rm F, eff}=t_2\sin(\pi n)$.
Note that $U_{\rm eff}/t_{\rm eff}$ can be large even when $U/t_1$ is small,
e.g., $U_{\rm eff}/t_{\rm eff}=8.8$ for $U=t_1$, $V=0$, and $t_2/t_1=0.1$.

For $V<V_c$ the quarter-filled dimerized extended
Hubbard model describes a Mott--Hubbard 
insulator with gap-less spin excitations.
Inn this parameter region, the spin degrees of freedom of 
the effective single-band Hubbard model~(\ref{hameffext}) can be described
by an effective Heisenberg model,
\begin{equation}
\hat{H}_{\rm heis, eff} = J_{\rm eff} \sum_{l_{\rm d}} 
\hat{{\bf S}}_{l_{\rm d}} \cdot \hat{{\bf S}}_{l_{\rm d}+1},
\label{effectiveHeisenberg}
\end{equation}
where $\hat{{\bf S}}_{l_{\rm d}}$ 
is the spin operator for a dimer located at position $l_{\rm d}$.
Up to second-order in $t_2/U_{\rm eff}$, we have
\begin{equation}
J_{\rm eff}(V) = \frac{4t_2^2}{8t_1+2U+V-2\sqrt{(U-V)^2+16t_1^2}} \; .
\label{fullJeff}
\end{equation}

\section{RESULTS}
\label{sec:results}

\subsection{Dimerized Hubbard model}

First, we consider the dimerized Hubbard model, i.e., we set $V=0$ 
in~(\ref{ham}).

\subsubsection{Tomonaga--Luttinger parameter}

In order to demonstrate the accuracy of our method, we address the
Tomonaga--Luttinger parameter at small interactions, $U<W$, as
a function of the dimerization in the metallic regime, $n=0.4$.
To lowest order in the couplings, $g_1=g_2=g_4=U/2$,
the field-theoretical `$g$-ology' approach predicts~\cite{Solyom,Giamarchibuch}
\begin{equation}
K_{\rho}=\sqrt{\frac{2\pi v_{\rm F}}{2\pi v_{\rm F}+U}}
\label{Krhog}
\end{equation}
where the Fermi velocity $v_{\rm F}$ is given by 
\begin{equation}
v_{\rm F}=
\frac{t_1t_2\sin k_{\rm F}}{\sqrt{t_1^2+t_2^2+2t_1t_2\cos(k_{\rm F})}}\; .
\end{equation}
This result can be systematically improved with the functional 
Renormalization Group method~\cite{fRG}.

\begin{figure}[t]
 \includegraphics[width=7.5cm]{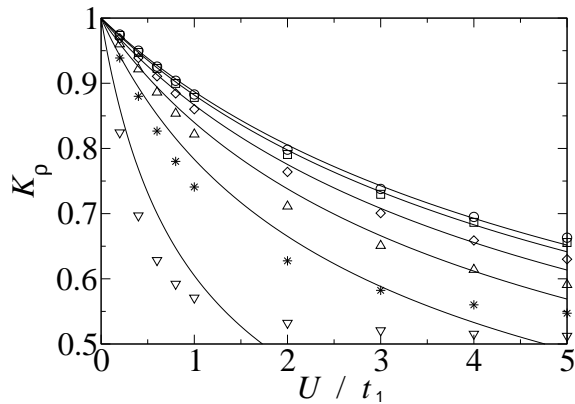}
\caption{Tomonaga--Luttinger parameter $K_{\rho}$ from the DMRG approach
(symbols) in comparison with the predictions from the $g$-ology method 
(solid lines), as a function of $U/t_1$ for $t_2/t_1 = 1, 0.9, 0.5, 0.3$,
$0.1$ (from top to bottom) at $n = 0.4$ for the dimerized Hubbard model.
\label{fig1}}
\end{figure}

In Fig.~\ref{fig1}, we compare the Tomonaga--Luttinger parameter 
as calculated from the DMRG approach, eq.~(\ref{KrhoD}), 
to the $g$-ology prediction~(\ref{Krhog}). We plot~$K_\rho$ 
as a function of $U/t_1$ for various dimerization 
strengths $t_2/t_1$ at band filling $n=0.4$. 
The system is metallic for all interaction strengths. 
For all dimerizations, $K_{\rho}$ decreases monotonically with increasing 
Coulomb interaction and finally approaches $K_{\rho}(U\to \infty)=1/2$,
as expected from the nondimerized Hubbard model.
For small dimerization,  $t_2/t_1\gtrsim 0.5$, the DMRG results agree 
very well with those from the $g$-ology approach for all~$U<W$.
For small $U/t_1$, $K_\rho$ decreases weakly and
monotonically with $t_2/t_1$. This can be understood from the corresponding
decrease of the bandwidth, $W=2(t_1+t_2)$, with a corresponding reduction 
of the Fermi velocity.

When the dimerization is large, $t_2/t_1 \lesssim 0.5$, and the
Hubbard interaction is large, $U \gtrsim W/2$, 
the results from $g$-ology substantially deviate from the numerically exact
DMRG results. The Tomonaga--Luttinger parameter $K_{\rho}$ decreases 
rapidly with decreasing $t_2/t_1$, and the $g$-ology predictions quickly
violate the constraint $K_{\rho}\ge 1/2$. Apparently, higher-order corrections
in $U/W$ beyond the one-loop calculations needed to be considered.

\begin{figure}[ht]
\includegraphics[width= 7cm]{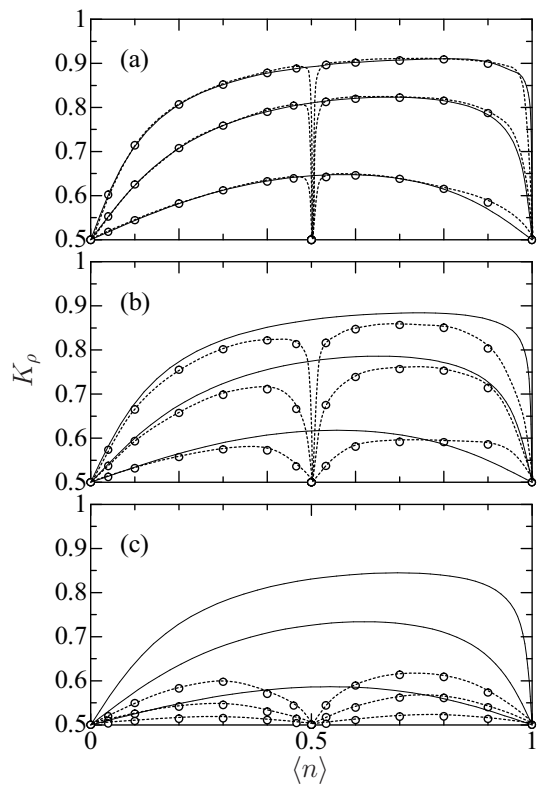}
  \caption{Tomonaga--Luttinger parameter $K_{\rho}$ 
as a function of the band-filling $n$ 
for various dimerizations:
(a)~$t_2/t_1=0.9$, (b)~$t_2/t_1=0.5$, and 
(c)~$t_2/t_1=0.1$. 
In each figure, $U/t_1=1,2,6$ from top to bottom. Open 
circles denote the DMRG results in the dimerized Hubbard model, 
and dotted lines are guides for eyes. Solid lines give
the exact result for the single-band Hubbard model with hopping integral 
$t=(t_1+t_2)/2$.\label{fig2}}
\end{figure}

As our next application, we investigate the Tomonaga--Luttinger parameter 
as a function of the band filling and the interaction strength. 
In Fig.~\ref{fig2}, we show $K_{\rho}$ from the DMRG method 
as a function of~$n$ for various interaction strengths $U/t_1$ 
and dimerizations: (a)~$t_2/t_1=0.9$, (b)~$t_2/t_1=0.5$, and 
(c)~$t_2/t_1=0.1$. 
For comparison we also plot the exact results for $K_{\rho}$ 
from the Bethe Ansatz for the one-dimensional single-band Hubbard model
with the same band width, $t=(t_1+t_2)/2$. 

When the dimerization is small, $t_2/t_1=0.9$,
we again find a good general agreement 
between the results for the dimerized Hubbard model and the single-band
Hubbard model with the same total bandwidth. 
An exception is the 
narrow range around quarter band filling, $n = 1/2$. 
At quarter filling, the lower Peierls band is half filled and 
the Umklapp scattering becomes a (marginally) relevant perturbation which
turns the metallic phase into a Mott--Hubbard insulator where 
$K_{\rho}$ is not well defined, and we give the value
for infinitesimal doping, see~eq.~(\ref{Krhoquarter}).
As expected from field theory~\cite{giamarchi,Giamarchibuch}, 
and confirmed numerically, we have
\begin{equation}
K_{\rho}\left(n=\frac{1}{2}^{\pm}\right)=\frac{1}{2}
\end{equation}
for the density-driven Mott transition for {\sl all\/} interaction strengths.
This follows from the mapping of the quarter-filled dimerized
Hubbard model to the effective single-band Hubbard model 
at half band-filling.
Therefore, $K_{\rho}$ strongly changes as a function of density in the vicinity
of quarter filling even for small dimerizations.
The effect becomes more prominent with increasing dimerization strengths, 
see Fig.~\ref{fig2}b.

When the dimerization is large, $t_2/t_1=0.1$, the single-band Hubbard model
does not provide a good starting point for the analysis anymore. 
Instead, for large $t_1/t_2$ we rather consider the effective
single-band Hubbard model~(\ref{hameffext}) for $V=0$.
Because of the strong effective on-site interaction
$U_{\rm eff}/t_{\rm eff}$, the Umklapp scattering strength becomes very 
large. For instance, the effective couplings at $t_2/t_1=0.1$ are
estimated from eq.~(\ref{effUext}) as 
$U_{\rm eff}/t_{\rm eff}=8.8, 15.3, 27.9$ 
for $V=0$ and $U/t_1=1,2,6$, respectively. Therefore, the values for $K_{\rho}$
are rather small for all $U/t_1\gtrsim 1$. Moreover, the effective single-band
Hubbard model always gives the correct result
$K_{\rho}(n=1/2^{\pm})=1/2$ because the quarter-filled
dimerized Hubbard model maps onto the half-filled single-band Hubbard model
which describes a Mott--Hubbard insulator for all interaction strengths.

\begin{figure}[ht]
\includegraphics[width= 7.5cm]{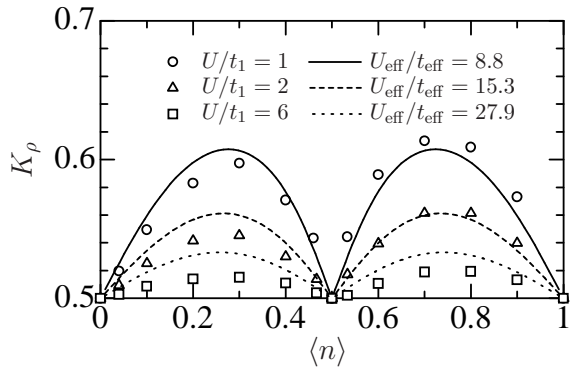}
  \caption{Tomonaga--Luttinger parameter $K_\rho$ from the DMRG approach
for large dimerization, $t_2/t_1=0.1$, in comparison with
the analytical result for the effective single-band Hubbard model.
Recall that the band filling $n$ of the dimerized Hubbard model
corresponds to a filling $2n$ for the effective single-band Hubbard model.
\label{fig3}}
\end{figure}

As seen in Fig.~\ref{fig3}, the quantitative agreement for $K_{\rho}$
from the dimerized Hubbard model and the effective single-band Hubbard
model is quite good for all $U/t_1$ at $t_2/t_1=0.1$. Note that
the effective Hubbard model displays its particle-hole symmetry around $n=1/2$
which the dimerized Hubbard model obeys only for $t_2/t_1\to 0$
or $U/t_1 \to \infty$.

\subsubsection{Spin excitations}
\label{subsubsect:spinex}

\begin{figure}[ht]
 \includegraphics[width= 7.5cm]{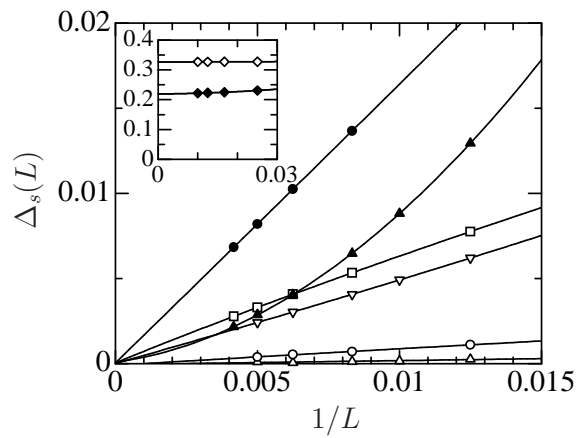}
\caption{Extrapolation of the spin gap $\Delta_s(L)$ of the
dimerized Hubbard model. Solid symbols represent the results 
for weak dimerization ($t_2/t_1=0.9$, $U/t_1=5$) 
at infinitesimal doping of the band-Mott insulator
($n=1$, triangles) and at small doping ($n=0.95$, circles).
Open symbols give the results for intermediate dimerization
($t_2/t_1=0.5$, $U/t_1=10$) at infinitesimal 
doping of the band-Mott insulator
($n=1$, triangles), at the electron densities 
$n=0.95$ (circles), $n=0.8$ (squares), and 
at infinitesimal doping of the Mott--Hubbard insulator at quarter filling
($n=0.5$, lower triangles). \\
Inset: Extrapolation of the spin gap of the band-Mott insulator 
at half band-filling for ($t_2/t_1=0.9$, $U/t_1=5$) 
(solid diamonds) and for ($t_2/t_1=0.5$, $U/t_1=10$) (open diamonds).
\label{fig4}}
\end{figure}

As our second quantity of interest we study the spin degrees of freedom 
at and around some commensurate band fillings. At half 
filling, $n = 1$, the dimerized Hubbard model is a band-Mott insulator
for all $U/t_1>0$, and we expect 
a finite gap for spin excitations for all $U/t_1$.
For small interaction strengths, the spin gap is of the order of
the Peierls gap, $\Delta_s(U/t_1\to 0)=\Delta_{\rm P}=2(t_1-t_2)$. 
For large interactions,
the spin degrees of freedom of the dimerized Hubbard model
can be described by the one-dimensional Peierls--Hubbard model
so that the spin gap to lowest order in $t_1/U$ becomes
\begin{equation}
\Delta_s(t_1/U\to 0) \propto \frac{4t_1^2}{U}
\left(\frac{t_1^2-t_2^2}{t_1^2+t_2^2}\right)^{2/3} \; ,
\label{spingap-n-1}
\end{equation}
in accordance with the results for the corresponding 
Peierls--Heisenberg model~\cite{bray}. Eq.~(\ref{spingap-n-1}) is applicable
for $U/t_1 \gtrsim 4$. In the inset of Fig.~\ref{fig4} we show
two examples for the finite-size scaling 
of the spin gap~(\ref{define-spingap}), 
$(t_2/t_1=0.5,U/t_1=10)$ and  $(t_2/t_1=0.9,U/t_1=5)$.
The dependence of the gap on the system size is quite small
because in the ground state individual spin singlets are formed on the dimers
so that the gapped spin excitations are rather localized in space.

It is more interesting to study the doping dependence of the spin gap.
In Fig.~\ref{fig4} we plot $\Delta_s(L)$ as a function
of system size for ($t_2/t_1=0.5,U/t_1=10$) and 
for ($t_2/t_1=0.9,U/t_1=5$) for several band fillings.
As seen from the figure, the spin gap vanishes for
all electron densities. In particular, at half band-filling 
it disappears as soon as the system is doped with an infinitesimal 
amount of holes. This can be understood in terms of the spin excitations
of a half-filled system with two holes.
Let us assume that the two holes are confined to a dimer.
Then, a spin excitations would remain the same local excitation as
in the perfectly half-filled system which costs the finite 
energy~(\ref{spingap-n-1}).
However, the holes are actually delocalized over the system because
the breaking of two spin dimers cost twice~$\Delta_s$ but the gain
in kinetic energy is approximately
\begin{equation}
E_{\rm it} \simeq 2(t_1-t_2) \; .
\label{eit}
\end{equation}
$E_{\rm it}$ is always larger than $2\Delta_s$. The mobile holes leave behind
at least two broken spin dimers whose spin excitation energy vanishes
in the thermodynamic limit. 

Apparently, the dimerized Hubbard behaves differently from 
the two-leg Hubbard ladder at half band-filling 
where a spin-singlet pair is formed 
on each rung. The spin gap in the ladder system remains finite 
for finite hole doping. There the spin-singlet pairs themselves are mobile
so that in the ground state 
an additional pair of holes is actually confined to a rung because 
the gain in kinetic energy due to the hole motion
is smaller than the combined loss in the pairing energy and
the kinetic energy of the spin dimers.

Finally, we investigate the spin gap for the quarter-filled
dimerized Hubbard model at infinitesimal doping.
In Fig.~\ref{fig4} we plot the size-dependence of the spin gap
for the infinitesimally doped Mott--Hubbard insulator at quarter filling
for ($t_2/t_1=0.5, U/t_1=10$). 
The extrapolated values are zero for all dimerization 
and interaction strengths. Therefore, the spin-gap liquid, 
suggested in the one-dimensional dimerized $t$-$J$ model~\cite{nishim2} 
is not realized in the dimerized Hubbard model. 

\subsection{Dimerized extended Hubbard model}

Now we turn to the case $V\neq 0$ in~(\ref{ham}). We focus on
the region around quarter filling where the nearest-neighbor interaction
can lead to a CDW phase.
This is known for the extended Hubbard model whose
ground-state phase diagram was studied in detail recently~\cite{mila,ejima}.

\subsubsection{Charge order}
\label{subsubsec:CO}

\begin{figure}[t]
    \includegraphics[width= 7.5cm]{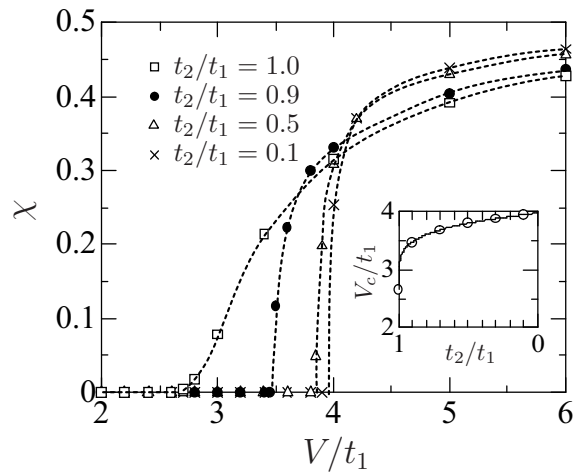}
  \caption{CDW order parameter $\chi$ 
extrapolated to the thermodynamic limit $L \to \infty$ 
for $t_2/t_1=1$, $0.9$, $0.5$, and $0.1$ with fixing $U/t_1=10$ 
at quarter filling. Lines are guides to the eyes. 
Inset: Estimated critical interaction strength
$V_c/t_1$ for the CDW transition 
as a function of $t_2/t_1$.\label{fig5}}
\end{figure}

Previous studies~\cite{tsuchiizu,shibata,clay3} suggested 
that the presence of a dimerization suppresses the CDW phase.
Therefore, we investigate the dependence of the critical coupling~$V_c$ for
the onset of the CDW. To this end we calculate
the CDW order parameter~$\chi$ from~(\ref{orderparameter})
as a function of the dimerization strength. 
For $V=0$ we have $\chi=0$ whereas, for  large $V$, 
the CDW order parameter approaches its classical value,
$\chi(V\to\infty)=0.5$.

In Fig.~\ref{fig5} we show the order parameter $\chi(V)$ as a function 
of~$V/t_1$ for dimerizations $t_2/t_1=1,0.9,0.5, 0.1$ 
for fixed $U/t_1=10$ at quarter band filling. 
In the absence of a Peierls modulation, $t_2/t_1=1$, i.e., in the extended
single-band Hubbard model, 
$\chi(V)$ is finite above $V_c/t_1 \approx 2.65$,
in agreement with previous work~\cite{mila,sano,ejima}.
Apparently, the dimerization enhances the charge fluctuation 
on each dimer, and, consequently the tendency towards charge order is 
reduced. 

In the presence of a dimerization the critical value for the onset
of the CDW increases with increasing dimerization.
Moreover, $\chi(V,t_2/t_1<1)$ rises up sharply 
above $V_c(t_2/t_1)$ even when $t_2/t_1$ is close to unity.
We speculate that the transition remains continuous for all finite $t_2/t_1$
but the slope is infinite for all $t_2/t_1>0$.
In the inset of Fig.~\ref{fig5} we show 
the critical value $V_c/t_1$ as a function of the dimerization 
strength $t_2/t_1$. We find that $V_c/t_1$ changes rapidly for small
$t_2/t_1$ and quickly saturates at its classical value for $t_2/t_1=0$.
The value $V_c(t_2/t_1=0)=4t_1$ is readily explained by considering 
an isolated dimer. In the isolated-dimer limit
the energies of the Mott--Hubbard 
insulator and the CDW insulator are 
\begin{eqnarray}
E_0^{\rm MH}/L &=& -t_1+V_{\rm eff}=-t_1+V/4 \; ,
\label{MHenergy} \\
E_0^{\rm CDW}/L&=&0 \; ,
\label{COenergy}
\end{eqnarray}
so that the criterion for the (discontinuous) transition
is $E_0^{\rm MH}(V_{c})=E_0^{\rm CDW}(V_{c})$ which immediately gives
$V_c/t_1 =4$. 

\subsubsection{Tomonaga--Luttinger parameter}

In the absence of a dimerization, 
the Tomonaga--Lut\-tinger parameter decreases 
as a function of $V/t$ for fixed $U/t>4$ and reaches $K_{\rho}=0.25$ at the 
critical coupling. When the CDW insulator 
is infinitesimally doped the system 
metalizes and $K_{\rho}^{\rm CDW}=1/8$~\cite{giamarchi,schulz,ejima}. 

For a finite dimerization, 
the quarter-filled system 
is a Mott--Hubbard insulator for small $V/t_1$ and finite $U/t_1$. 
At infinitesimal doping we find $K_{\rho}^{\rm MH}(V<V_c)=1/2$
below the transition, independent of~$V$.
This is readily understood from the fact that the effective model
is the extended single-band Hubbard model at half band-filling
for which the field-theoretical arguments for a density-driven
Mott transition still apply. A qualitatively and quantitatively 
different behavior emerges from the transition to
the CDW insulator at $V_c$.
The Tomonaga--Luttinger parameter drops from $K_{\rho}=1/2$
in the infinitesimally doped Mott--Hubbard insulator to
$K_{\rho}<1/8$, as we shall discuss in more detail now.

The dimerization has two prominent effects on $K_{\rho}$.
First, it increases the strength of the Umklapp scattering 
which makes $K_{\rho}$ smaller. Second, the dimerization suppresses the 
CDW instability which tends to make $K_{\rho}$ larger. 
These effects are most apparent around quarter filling where
the two tendencies compete with each other close to the CDW instability. 
Both effects increase upon decreasing $t_2/t_1$. The first effect continues to 
develop progressively and leads to $U_{\rm eff}/t_{\rm eff} \to \infty$ 
as $t_2/t_1 \to 0$. As shown in Sect.~\ref{subsubsec:CO}, 
the second effect develops fast as a function of the dimerization
 and quickly saturates.
Therefore, we expect that the first effect, a reduction of $K_{\rho}$
upon dimerization, is more prominent but for quarter filling and in the
vicinity of the transition to the CDW phase.

The reduction of $K_{\rho}$ with dimerization 
can actually be inferred from the $g$-ology approach where 
the Tomonaga-Luttinger parameter near quarter filling is given by
\begin{eqnarray}
K_{\rho} \approx \sqrt{\frac{2\pi v_{\rm F}-V}{2\pi v_{\rm F}+U+5V}}
\label{KrhogwithV}
\end{eqnarray}
with $v_{\rm F}=t_1t_2/\sqrt{t_1^2+t_2^2}$. 
The formula shows that $K_{\rho}$ decreases monotonously 
as a function of $V$ and of $t_2/t_1$.
Naturally, $g$-ology cannot cover large dimerizations or 
the transition region where 
the increase of $K_{\rho}(V)$ upon dimerization becomes apparent.

\begin{figure}[t]
\includegraphics[width= 7.8cm]{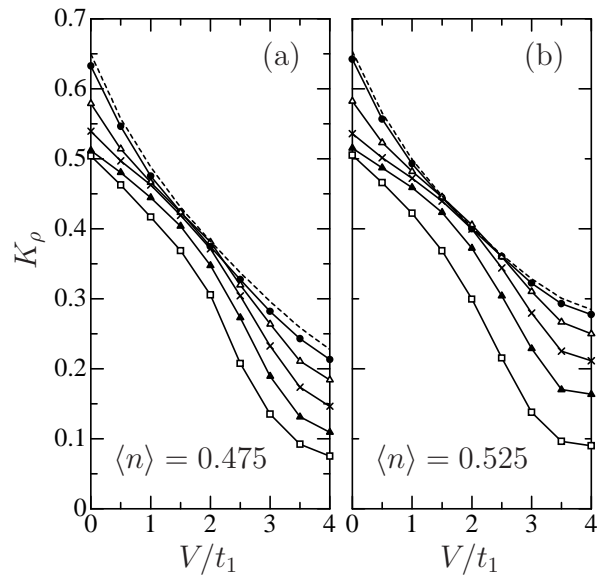}
\caption{Tomonaga--Luttinger parameter $K_{\rho}$ in the dimerized extended 
Hubbard model as a function of the nearest-neighbor Coulomb interaction 
$V/t_1$ for $U/t_1=6$ and various dimerizations: 
$t_2/t_1=1$ (dashed line),
$t_2/t_1=0.9$ (filled circles), $t_2/t_1=0.7$ (open triangles), $t_2/t_1=0.5$ 
(crosses), $t_2/t_1=0.3$ (filled triangles), and $t_2/t_1=0.1$ (open squares).
The band filling is (a)~$n = 0.475$ 
and (b)~$n = 0.525$.\label{fig6}}
\end{figure}

In Fig.~\ref{fig6} we present the DMRG results for $K_{\rho}$ as a function
of $V/t_1$ at a hole doping and an electron doping of 5\%,
$\langle n \rangle = 0.5 \pm 0.025$, for $U/t_1=6$ and various dimerizations.
The numerically exact DMRG results confirm the general expectations
as expressed in the $g$-ology formula~(\ref{KrhogwithV}).
The Tomonaga--Luttinger parameter decreases monotonously
with~$V/t_1$ for all dimerizations and, in general, 
it decreases as a function
of $t_2/t_1$ for fixed~$V/t_1$. The fact that $K_{\rho}$ is almost
independent of $t_2/t_1$ for fixed $1<V/t_1<2$ can be attributed 
to the above-mentioned competition
between the Umklapp scattering and the charge ordering.
For certain parameter regions, a change in the dimerization strength
has almost no net effect on $K_{\rho}$ because a change in the strength of the
Umklapp scattering is compensated by a change in the mobility
of the charge carriers.
For the same parameter set $(V/t_1,t_2/t_1)$,
$K_{\rho}$ is generally somewhat smaller for the hole-doped case than for the
electron-doped case but there is no difference in the qualitative
behavior. This had to be expected because the system is particle-hole 
symmetric around quarter filling to lowest order in $t_2/t_1$. From now on
we shall focus on the case of hole doping.

The Tomonaga--Luttinger parameter~$K_{\rho}(V)$ changes most rapidly
in the region $2<V/t_1<4$ where the quarter-filled system undergoes 
the charge-ordering transition. 
For $V/t_1\gtrsim 4$, we can interpret the system as a doped 
CDW insulator. In this region, 
we find that the dependence of $K_{\rho}$ on the nearest-neighbor
interaction~$V/t_1$ is much weaker. This can be understood from  
the Taylor expansion of $K_{\rho}$ for a slightly doped
CDW insulator. Above the transition point ($V > V_c$) 
we generally expect~\cite{ejima} that for
$\delta = 1/2 -n \ll 1$ we have 
\begin{equation}
K_{\rho}(t_2,U,V,1/2-\delta) = K_{\rho}^{\rm CDW}(t_2,V) 
+ \frac{\delta}{h(t_2,U,V)} + \cdots \; ,
\end{equation} 
where $t_1$ is used as energy unit. 
The prefactor $h(t_2,U,V)$ diverges exponentially at
the critical interaction strength $V_c$ but it rapidly tends to a constant 
for large $V$. 

\begin{figure}[t]
\includegraphics[width=6.5cm]{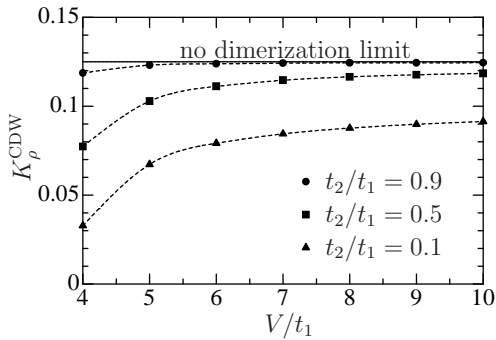}
\caption{Tomonaga--Luttinger parameter for the infinitesimally doped 
CDW insulator $K^{\rm CDW}_{\rho}$ 
as a function of $V/t_1$ for $t_2/t_1=0.9$, $t_2/t_1=0.5$, and $t_2/t_1=0.1$ 
at $U/t_1=\infty$. The solid line corresponds to $K^{\rm CDW}_{\rho}=1/8$ 
when the dimerization is absent ($t_1=t_2$), and the
dotted lines are guides for the eyes.\label{fig7a}}
\end{figure}

For infinitesimal doping,
the Tomonaga--Luttinger parameter of the CDW insulator
$K^{\rm CDW}_{\rho}(t_2/t_1,V/t_1)$ 
also displays a smooth behavior 
as a function of $V/t_1$ and $t_2/t_1$.
In Fig.~\ref{fig7a} we show $K^{\rm CDW}_{\rho}(t_2/t_1,V/t_1)$ 
for $U=\infty$ and various dimerizations. 
As for the case of a finite doping we see that
the dimerization tends to reduce the Tomonaga--Luttinger parameter.
In the CDW phase this tendency is somewhat compensated by the
influence of the nearest-neighbor Coulomb interaction which, for
large interactions and for small doping of the CDW state, 
delocalizes the holes over the system and therefore increases the
charge fluctuations which determine~$K_{\rho}$ via eq.~(\ref{eqn:den1}).

The most important observation is the magnitude of the
Tomonaga--Luttinger parameter for the doped insulators.
For infinitesimal doping we find $K_{\rho}=1/8$ in the absence
of dimerization and even $K_{\rho}<1/8$ in the presence of 
a dimerization. These small numbers persist for finite doping,
as seen in Fig.~\ref{fig6}. Therefore, depending on
the choice of the dimerization and the nearest-neighbor
Coulomb interaction, one can easily find parameter regions
where $0.1< K_{\rho} < 0.3$ can be realized for
slightly doped quarter-filled chains.

\begin{figure}[h]
\includegraphics[width= 7.0cm]{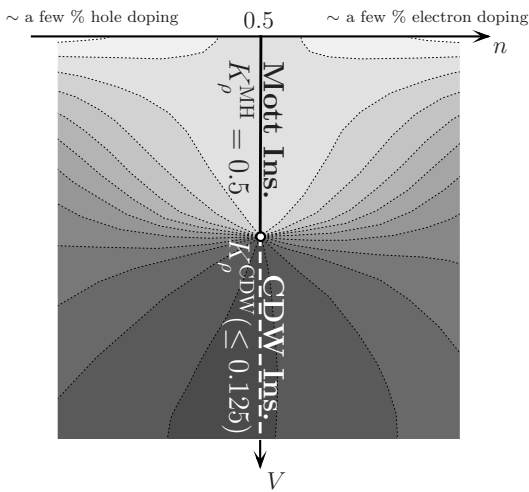}
 \caption{Schematic phase diagram of the dimerized extended Hubbard model 
around quarter filling as a function of $n$ and $V$. 
Variations of $K_{\rho}$ are displayed by contour lines. 
Darker (brighter) color denotes smaller 
(larger) values of $K_{\rho}$.\label{fig7b}}
\end{figure}

Fig.~\ref{fig7b} summarizes our findings for the Tomonaga--Luttinger
parameter in a schematic phase diagram 
for the slightly doped quarter-filled dimerized extended Hubbard model.
The Mott--Hubbard insulator (CDW insulator) 
can be characterized as $2k_{\rm F}$-SDW 
($4k_{\rm F}$-CDW) states at quarter filling. 
Hence, the $2k_{\rm F}$-SDW and $4k_{\rm F}$-CDW 
correlations are dominant for slightly doped Mott--Hubbard and CDW insulators,
respectively, and their correlation functions decay
algebraically with 
the asymptotical behavior $C^{2k_{\rm F}{\text -}{\rm SDW}}(r) \sim r^{-1-K_{\rho}}$ 
for Mott--Hubbard
insulators and $C^{4k_{\rm F}{\text -}{\rm CDW}}(r) \sim r^{-4K_{\rho}}$ for CDW insulators.
Thus, the value $K_{\rho}=1/3$ discriminates the two types of (doped)
insulators at finite doping.
For an infinitesimally doped insulator we correctly find
$K_{\rho}(V=V_{\rm dc})=1/3$ for $V_{\rm dc}=V_c$ but
$V_{\rm dc}$ becomes actually 
smaller upon doping, as seen in Fig.~\ref{fig7b}.

\subsection{Comparison with experiment}

At last, we compare our theoretical result with experiments on (TMTTF)$_2$X.
The electron transfer matrix elements~\cite{ducasse} 
are estimated to be $(t_1,t_2)=(137\, {\rm meV},93\, {\rm meV})$ for X=PF$_6$, 
$(t_1,t_2)=(140\, {\rm meV},100\, {\rm meV})$ for X=ClO$_4$, and 
$(t_1,t_2)=(133\, {\rm meV}, 119\, {\rm meV})$ for X=Br, i.e., 
$t_2/t_1= 0.68, 0.71, 0.89$ for X=PF$_6$, ClO$_4$, Br, respectively. From 
the comparison with the optical gap~\cite{milacond,benthien}
the Coulomb parameters are estimated to be $U/t_1 \approx 7.0$ and 
$V/t_1\approx 2.8$ for (TMTTF)$_2$PF$_6$. A comparison
with Fig.~\ref{fig6} and~\ref{fig7b} shows that this parameter set 
leads to $K_{\rho}\approx 0.25$, in agreement
with experimental estimates for the Tomonaga--Luttinger parameter
from the temperature dependence of the resistivity~\cite{moser,korin-hamzic}.
In view of the CDW state observed below 
$T \approx 100\, {\rm K}$~\cite{nad,chow,zamborszky}, the nearest-neighbor
interaction could be even larger than $V/t_1=2.8$ which would
further reduce $K_{\rho}$.

Unfortunately, such values for the nearest-neighbor interaction~$V/t_1$
appear to contradict the results for the effective exchange 
interaction as deduced from the high-temperature
data from the electron-spin-resonance (ESR) 
measurements~\cite{dumm}, 
$J_{\rm exp}=420\, {\rm K}, 430\, {\rm K}, 500\, {\rm K}$ for
the anions X=PF$_6$, ClO$_4$, Br, respectively.
In the presence of the dimerization and at quarter band-filling
we can start from the effective extended
single-band Hubbard model~(\ref{hameffext}) and the
spin degrees of freedom can be described in terms of
the effective Heisenberg Hamiltonian~(\ref{effectiveHeisenberg}).
For $U/t_1=7.0$, the bare Hubbard model, $V=0$ in~(\ref{fullJeff}), gives
$J_{\rm eff}(V=0)=499\, {\rm K}, 564\, {\rm K}, 841\, {\rm K}$.
The good agreement of the experimental and theoretical data for $V=0$
implies that the nearest-neighbor interaction ought to be rather small. 
In particular, the a value $V=2.8 t_1$ for 
(TMTTF)$_2$PF$_6$, leads to $J_{\rm eff}(V=2.8 t_1)=222\, {\rm K}$,
a factor of two smaller than the experimental estimate.
Additinally, with small $V$ to adjust $J_{\rm exp}$,  
the resulting theoretical prediction for $K_{\rho}\approx 0.5$
from Fig.~\ref{fig6} is not compatible 
with the experimental estimate, $0.2 \lesssim K_{\rho} \lesssim 0.3$.

In order to reconcile this discrepancy we note that, in the ESR measurements, 
the curves are fitted to provide a good agreement with the 
Eggert--Affleck--Takahashi model~\cite{eggert} for the spin susceptibility
of the $S=1/2$ antiferromagnetic Heisenberg chain 
at elevated temperatures. However, substantial deviations occur
for small temperatures, $T \lesssim 100\, {\rm K}$.
They could be the result of a dimensional crossover~\cite{dressel2} and the
transition to the CDW phase. We
are tempted to attribute the deviations to an effectively larger
nearest-neighbor interaction at low temperatures. 
Recall that our electronic model is purely one-dimensional,
and neither covers the influence of phonons~\cite{clay1,maurel} 
nor does it give an
account on the screening of the electron-electron interaction
which may change drastically in the vicinity of the transition
to the CDW state.
Therefore, temperature may have a quite substantial influence 
on the value of the effective $V$-parameter in our model
so that eq.~(\ref{fullJeff}) cannot be applied with the values
for $V/t_1$ at $T=0$ to explain the susceptibility data for $T>100 \, {\rm K}$.

In fact, in the CDW phase, the effective exchange
interaction is given by
$J^{\rm CDW}_{\rm eff}/t_1 \approx 4t_2^4/(2UV^2)$ which results
in $J^{\rm CDW}_{\rm eff}= 14\, {\rm K}$ if we use the parameters
for (TMTTF)$_2$PF$_6$. If the spin susceptibility could be measured
in the (one-dimensional)
CDW phase, the exchange interaction should be an order
of magnitude smaller than in the high-temperature phase.
 
In (TMTSF)$_2$PF$_6$, the hopping amplitudes are estimated
as $(t_1=252\, {\rm, meV}, t_2=209\, {\rm meV})$ and  
the effective Coulomb interactions are found to be weaker, 
$U/t_1 \sim 5$. Again, a weak nearest-neighbor 
Coulomb interaction, $V_1\approx 0.5 t_1$, would account for an 
exchange interaction $J_{\rm eff}=1.2 \cdot 10^3 \, {\rm K}$ 
which is compatible
with the high-temperature experimental observation 
$J_{\rm exp}\approx 1.4 \cdot 10^3 \, {\rm K}$.

\section{SUMMARY}
\label{sec:sum}

Using the DMRG method, we provided numerically exact
results for the spin excitations, the CDW order parameter,
and the Tomonaga--Luttinger parameter
of the one-dimensional dimerized extended Hubbard model 
at and near commensurate fillings.

In the presence of a dimerization we confirm numerically
that gap for the spin excitation is finite at half band-filling.
However, the gap immediately disappears when the system is doped 
infinitesimally because there is no mechanism which confines the holes
to a single dimer. This result is qualitatively consistent with 
a rapid suppression of the spin gap with Zn doping 
in the spin-Peierls Heisenberg system CuGeO$_3$~\cite{oseroff}, 
irrespective of the difficulty in metalization this material~\cite{terasaki}.

For the Tomonaga--Luttinger parameter the effects of the dimerization 
are weak in the absence of the nearest-neighbor Coulomb interaction~$V$ 
and away from quarter filling.
At and near quarter filling, the lower Peierls band is essentially half
filled and the dimerized Hubbard model at filling $n=1/2\pm \delta$
can be understood qualitatively and even semi-quantitatively 
in terms of an effective single-band Hubbard model at electron 
density~$2n$. From the result of the corresponding Hubbard model 
at half band-filling it immediately follows that
$K_{\rho}=1/2$ holds for the dimerized Hubbard model
at infinitesimally doping away from quarter filling.
Therefore, the Tomonaga--Luttinger parameter
for the weakly doped quarter-filled 
system sensitively depends on the strength of the dimerization.
In general, the dimerization tends to reduce $K_{\rho}$ gradually
because the effective scattering processes within the Peierls bands increase
with the size of the Peierls gap.

In the presence of the nearest-neighbor Coulomb interaction,
the case of quarter filling also deserves special attention because
the Mott--Hubbard insulator goes over to a CDW insulator
with a finite spin gap at a critical interaction strength~$V_c$.
The dimerization opposes the formation of the CDW phase,
for example, the critical nearest-neighbor interaction
shifts from $V_c/t_1\approx 2.65$ in the absence of dimerization to
$V_c/t_1=4$ in the dimer limit.

The suppression of the charge order at quarter filling by the dimerization
is reflected in a tendency to stabilize the metallic state by the dimerization
away from quarter band-filling. However, the increase of the electron-electron
scattering by the nearest-neighbor Coulomb 
interaction overcomes that tendency and results 
in a net reduction of $K_{\rho}$ as a function
of the dimerization and the nearest-neighbor interaction, see Fig.~\ref{fig6}.
As a consequence, fairly small values, $K_{\rho}\approx 0.25$, 
can be obtained for a moderate five-percent 
doping of the quarter-filled dimerized extended Hubbard model 
at moderate Coulomb couplings, $U/t_1=6$, $V/t_1=3$.

It is difficult to reconcile all experimental data for the Bechgaard salts
with our findings for the dimerized extended Hubbard model in one dimension.
In order to find small values for the Tomonaga--Luttinger parameter,
the Coulomb interactions must be large enough to reach the region
of a (doped) CDW insulator which is not easily
reconciled with the high-temperature data for the exchange interaction.
We suspect that the one-dimensional dimerized extended Hubbard model  
is still too simplistic to describe the physics
of the Bechgaard salts adequately.

\acknowledgments
We thank E.\ Jeckelmann for useful discussions.
S.E.\ is supported by the Honjo 
International Scholarship Foundation.

\end{document}